\documentclass[reqno]{article}%12pt%
\usepackage{alltt}
\usepackage{graphicx}
\title{Super-relativity in the quantum theory}
\author{Peter Leifer}
\date{ Cathedra of Informatics, Crimea State Engineering and
Pedagogical University, \\
21 Sevastopolskaya st., 95015 Simferopol, Crimea, Ukraine }
\begin{document}
\maketitle
\begin{abstract}
The relativity to the measuring device in quantum theory, i.e. the
covariance of local dynamical variables relative transformations to
moving quantum reference frame in Hilbert space, may be achieved
only by the rejection of super-selection rule. In order to avoid the
subjective nuance, I emphasis that the notion of ``measurement"
here, is nothing but the covariant differentiation procedure in the
functional quantum phase space $CP(N-1)$, having pure objective
sense of evolution. Transition to the local moving quantum reference
frame leads to some particle-like solutions of quasi-linear field
PDE in the dynamical space-time. Thereby, the functionally covariant
quantum dynamics gives the perspective to unify the Einstein
relativity and quantum principles which are obviously contradictable
under the standard approaches.
\end{abstract}


\section{Introduction}
There is some analogy between classical accelerated reference frame
in general relativity and moving quantum reference frame in Hilbert
space. From the formal point of view in both cases arise physical
fields of quite different nature. It would be interesting to find a
geometric way to their unification.

The backreaction of the ``light'' sub-system on the ``heavy''
sub-system dynamics through a gauge vector and scalar potentials was
already geometrized \cite{Berry187}. It comes from the adiabatic
Born-Oppenheimer ansatz applied to the coupled sub-systems. On the
other hand Aharonov {\it et al} made accent on the quantum reference
frame - ``second particle'' reaction described by the effective
topological vector potential due to the measurement process
\cite{AA88,A97}. Authors emphasize the necessity of ``second
particle'' usage. Cartan's method of the moving frame, however, rids
us from this obligation. Namely, the quantum state and dynamical
variables of a quantum system may be referred to the infinitesimally
close previous state. Thereby one is able to restore the objective
interpretation of the quantum description. But transition to moving
frame in Hilbert space requires the especial control of the elements
proximity since the non-adiabatic approximation (where transitions
are available) realized by transition to moving frame leads to
divergences of the iteration process even for two-level system
\cite{Berry164}.

The fundamental gauge field coming ``from nowhere" in the models of
elementary particles, and both Abelian \cite{Berry187} and
non-Abelian \cite{WZ} pseudo-potentials associated with adiabatic
Born-Oppenheimer approximation, have formally geometric origin.
Pseudo-potentials have a singular source of monopole-like type whose
nature arose under degeneration, etc. Dirac put the monopole as a
physical source of electromagnetic field. He assumed that
singularities are concentrated on some ``line of the knots" in the
physical space. But the mathematical artefact, say, singularity of
mapping cannon be a reason for physical phenomenon. On the other
hand monopoles do not exist as physical object up to now; one should
have physical mechanism of the fields generation.

The first question is: is it possible to find a non-singular
fundamental gauge potential if one uses local coordinates in
inherently related projective Hilbert space $CP(N-1)$ instead of the
parameter space?

The second question was posed by Berry: ``What is the dynamical
significance of the moving frame that produce the best approximation
to $\gamma$,...?" \cite{Berry187}.

There are two reasons of ``defeats" of the renormalization procedure
by the successive transitions to moving frame \cite{Berry164,
Berry187}. The first of them is lurked in the local
(state-dependent) character of embedding both the isotropy sub-group
$H=U(1)\times U(N-1)$ of some state vector and the coset
transformations $G/H=U(N)/U(1)\times U(N-1)=CP(N-1)$ into the
$G=U(N)$ group. The second reason is that the Berry's condition of
the ``parallel transport" $<\psi|\dot{\psi}>=0$ is not affine, i.e.
it is nor invariant relative Fubini-Study metric neither even linear
\cite{AS}.

In the framework of my model the reason of anholonomy is generated
by the curvature of the dynamical group manifold and its invariant
sub-manifold $CP(N-1)$. Such geometry is the true source of some
non-singular physical fields. The affine parallel transport agrees
with Fubibi-Study leads presumably to the ``best approximation" in
quantum dynamics. It is interesting that the similar problem and the
necessity to take into account the structure of the projective
Hilbert space arise in quite different physical problem.

For quantum states of a compound system whose parts obeys the
super-selection rule (atom, e.g.) we have Schr\"odinger or
Heizenberg equations of motion. The problem to find an approximate
dynamical states has a formally clear formulation but it seldom has
the physically acceptable solutions for quantum field system
\cite{Dirac1,Dirac2}. Divergences arising here are rooted into
infinite degrees of freedom, that was demonstrated in very simple
example \cite{Dirac1}. It is shown \cite{Le1} that the orthogonal
projection along of the vacuum  state i.e. the subtraction the
`longitudinal' component of the variation velocity of the state
$|\Psi(t)> $ is merely partly helpful. Let put $\eta_n$ to be
creation operator and $|\xi_n> = \eta_n |S> = \alpha_n |S > + |n)$
is deformed standard vector, so that $(n|S
>=0$, therefore, $<S |\xi_n> = <S | \eta_n |S > = \alpha_n <S |S >$
and, hence, $\alpha_n = \frac{<S | \eta_n |S >}{<S |S >}$. Now we
can express the `transversal' part of the standard vector
deformation representing in fact the tangent vector to $CP(\infty)$
\begin{eqnarray}{\label{1}}
|n) = |\xi_n> - \frac{<S | \xi_n>}{<S |S>}|S >,
\end{eqnarray}
with the Hermitian orthogonality to $<S|$ \cite{Arnold}. Let me
calculate only the `transversal' components of the
$|\Psi(t)>=\exp(\frac{i}{\hbar}\hat{H}t)|S>$ which I will define as
follows:
\begin{eqnarray}
|\Psi(t)) = |\Psi(t)> - \frac{<S | \Psi(t)>}{<S |S >}|S
> \cr =\exp(\frac{i}{\hbar}\hat{H}t)|S>-\frac{|S> <S|}{<S |S >}
\exp(\frac{i}{\hbar}\hat{H}t)|S>
\end{eqnarray}
Now I apply this definition to calculation of all orders of
$|\Psi(t)>$:
\begin{eqnarray}
|\Psi(t)_0) = |S> - \frac{<S | S>}{<S |S >}|S >=|S>-\frac{|S>
<S|}{<S |S
>}|S>=0,
\end{eqnarray}

\begin{eqnarray}
|\Psi(t)_1) = t(\hat{H}|S> - \frac{<S |\hat{H}| S>}{<S |S >}|S
>) \cr =t(\hat{H}|S>-\frac{|S> <S|}{<S |S
>}\hat{H}|S>)=t\hat{H}|S>,
\end{eqnarray}

\begin{eqnarray}
|\Psi(t)_2) = \frac{t^2}{2!}(\hat{H}^2|S> - \frac{<S |\hat{H}^2|
S>}{<S |S
>}|S>)
 \cr = \frac{t^2}{2!}(\hat{H}^2|S>-\frac{|S> <S|}{<S|S>}\hat{H}^2|S>)
=\frac{t^2}{2!}\frac{1}{4}a_{mn}a_{pq}\eta_m \eta_n \eta_p \eta_q
|S>,
\end{eqnarray}
where terms, giving nil contribution were omitted. We can see that
divergent second order term was canceled out. But in the third order
one has
\begin{eqnarray}
|\Psi(t)_3) = \frac{t^3}{3!}(\hat{H}^3|S> - \frac{<S |\hat{H}^3|
S>}{<S |S
>}|S>) \cr
=\frac{t^3}{3!}\frac{1}{8} (a_{mn}a_{pq}a_{rs}\eta_m \eta_n \eta_p
\eta_q \eta_p \eta_s-2 Tr(\bar{a}a) a_{mn}\eta_m \eta_n)|S>,
\end{eqnarray}
hence one sees that divergences alive in the third order and that
the compensation projective term does not help. Nevertheless, we can
extract the useful hint: the vacuum vector (the standard vector in
Dirac's example) should be smoothly changed, and, furthermore,  the
transversal component should be reduced during the ``smooth''
evolution. One may image some a smooth ``surface" in the functional
state space with a normal vector, taking the place of the vacuum
vector and tangent vectors comprising together the local reference
frame. It is natural to start with finite dimensional complex
projective Hilbert space $CP(N-1)$ closely related to eigen-problem
and where limit $N \to \infty$ may be easy achieved \cite{Le1,Le2}.
Then the orthogonal projection acting continuously is in fact the
complex covariant differentiation of the tangent vector fields
\cite{KN}.

The assumption that local moving reference frame in the complex
projective Hilbert space generates some fields was called
super-relativity \cite{Le2}. Intrinsic unification of quantum and
relativity and the verification of the physical status of these
fields require dynamical (state-dependent) space-time construction
and modification of the second quantization method.

\section{Dynamical state-dependent space-time}
The peaceful coexistence between quantum behavior and classical
relativity (both special and general) is highly desirable but it
looks like the far distant future \cite{Penrose}. All known attempts
to reach the harmony (strings, super-gravity, e.g.) lead to very
strange unobservable predictions (super-partners) and they require
entirely to change our scientific paradigm (anthropic principle,
Multiverse). I propose much more modest approach. It was recognized
that the space-time is macroscopically observable as global
pseudo-riemannian manifold is only obtrusive illusion from the
quantum point of view. Namely, the operational coordinatization of
classical events by means of {\it electromagnetic field} is based on
the distinguishability, i.e. individualization of pointvise material
points. However we loss this possibility by means of {\it quantum
fields}. Generally, it is important to understand that the problem
of identification is the root problem even in classical physics and
that its recognition gave to Einstein the key to formalization of
the relativistic kinematics and dynamics. Indeed, only assuming the
possibility to detect locally the coincidence of two pointwise
events of a different nature it is possible to build all kinematic
scheme and the physical geometry of space-time
\cite{Einstein1,Einstein2}. As such the ``state" of the local clock
gives us local coordinates - the ``state" of the incoming train. In
the classical case the notions of the ``clock" and the ``train" are
intuitively clear. Furthermore, Einstein especially notes that he
did not discuss the inaccuracy of the simultaneity of two {\it
approximately coincided events} that should be overcame by some
abstraction \cite{Einstein1}. This abstraction is of course the
neglect of finite sizes (and all internal degrees of freedom) of the
both real clock and train. It gives the representation of these
``states" by mathematical points in space-time. Thereby the local
identification of two events is the formal source of the classical
relativistic theory. But in the quantum case such identification is
impossible since the localization of quantum particles is
state-dependent \cite{W,Heg,A98}. Hence the identification of
quantum events (transitions) requires a physically motivated
operational procedure with corresponding mathematical description.

Therefore it is inconsistent to start the development of the quantum
theory from the space-time symmetries because just the space-time
properties should be established in some approximation to internal
quantum dynamics, i.e. literally {\it a posteriori}. Namely, the
quantum measurement with help of the ``quantum question" leads
locally to the Lorentz transformations of its spinor components,
and, on the other hand, to dynamical (state-dependent) space-time
coordinatization. That is, instead of the representation of the
Poincare group in some extended Hilbert space, I used an ``inverse
representation" of the $SU(N)$ by solutions of relativistic
quasi-linear PDE in the dynamical space-time.
\section{The action quantization}
The concept of the ``elementary particles" cannot be applied to the
problem of identification of quantum events since we are lacking for
consistent QFT at the sub-atomic level. In the absence of
super-selection rule, the role of the most universal building blocks
might play the action amplitudes. I propose some discrete quantum
model based on the concept of the ``elementary quantum states" (EQS)
$|\hbar a>, \quad a=0,1,...$ \cite{Le3,Le4,Le5,Le6}. In the
framework of this model it is assumed that the Planck's hypothesis
should be literally applied to the action quantization. It is clear
that the state $|\hbar a>$ belongs to some sector $\{a\}: |\hbar a>
\in \{a\}, \quad a=0,1,...$ and there is the continuum of EQS's in
each sector (space-time/energy-momentum distribution splits them in
a ``zone").

The space-time representation of EQS's and their coherent
superposition is postponed on the dynamical stage as it is described
below. We shall construct non-linear field equations describing
energy (frequency) distribution between EQS's $|\hbar a>$, whose
soliton-like solution provides the quantization of the dynamical
variables. Presumably, the stationary processes are represented by
stable particles and quasi-stationary processes are represented by
unstable resonances.

Since the action in itself does not create gravity, it is possible
to create the linear superposition of $|\hbar a>=(a!)^{-1/2} ({\hat
\eta^+})^a|\hbar 0>$ constituting $SU(\infty)$ multiplete of the
Planck's action quanta operator $\hat{S}=\hbar {\hat \eta^+} {\hat
\eta}$ with the spectrum $S_a=\hbar a$ in the separable Hilbert
space $\cal{H}$. This superposition physically corresponds to the
complete amplitude of some quantum motion or a process.

Generally the coherent superposition
\begin{eqnarray}
|F>=\sum_{a=0}^{\infty} f^a| \hbar a>,
\end{eqnarray}
may represent of a ground state or a ``vacuum" of some quantum
system with the action operator
\begin{eqnarray}
\hat{S}=\hbar A({\hat \eta^+} {\hat \eta}).
\end{eqnarray}
Such vacuum is more general than so-called ``$\theta$-vacuum"
\begin{eqnarray}
|\theta>=\sum_{a=-\infty}^{\infty} e^{i\theta a}| \hbar a>.
\end{eqnarray}
The ``winding number" $a$ has here different sense as it was
mentioned above.

The action functional
\begin{eqnarray}\label{af}
S[|F>]=\frac{<F|\hat{S}|F>}{<F|F>},
\end{eqnarray}
has the eigen-value $S[|\hbar a>]=\hbar a$ on the eigen-vector
$|\hbar a>$ of the operator $\hbar A({\hat \eta^+} {\hat
\eta})=\hbar {\hat \eta^+} {\hat \eta}$. This deviates in general
from this value on superposed states $|F>$ and of course under
non-trivial choice of the $A(.,.)$-function: $\hat{S}=\hbar A({\hat
\eta^+} {\hat \eta}) \neq \hbar {\hat \eta^+} {\hat \eta}$. The
relative (local) vacuum of some problem is not necessarily the state
with minimal energy but sometimes it may be interpreted as a
extremal of the action functional of a classical (or ordinary
quantum) variational problem.

In fact only finite, say, $N$ elementary quantum states (EQS's)
($|\hbar 0>, |\hbar 1>,...,|\hbar (N-1)>$) may be involved in the
coherent superposition $|F>$. Then $\cal{H}=C^N$ and the ray space
$CP(\infty)$ will be restricted to finite dimensional $CP(N-1)$.
Hereafter we will use the indices as follows: $0\leq a,b \leq N$,
and $1\leq i,k,m,n,s \leq N-1$. Due to the discreteness of the our
model, the variation problem may be reduced simply to the
differentiation of the complex tensor fields.

\section{The $SU(N)$ local dynamical variables}
Let me assume that one deals with the eigen-problem for some action
operator $\hat{S}$ acting in ${\cal{H}}=C^N$. Its  an eigen-vector
lies in the ray $Z|G>, \quad Z \in C$ and the last one marks the
extremum point of the action functional $S[|F>]$ belonging to
$CP(N-1)$. Since we have not yet the space-time representations for
this state and for any dynamical variables one should find the
invariant conditions for the stability of the extremum from the
$SU(N)$ geometry.

The unitary transformation $\hat{U}=\exp (i \hat{S})$ leaves this
ray intact $\hat{U}Z|G>=\exp (i \hat{S})Z|G>=Z\exp ( i \lambda )
|G>$. In order to see what happens in the case of general unitary
transformation applied to the eigen-vector, it is useful to use some
representation where $Z|G>$ has only one non-zero component, say,
\begin{eqnarray}\label{vac}
Z|G>= Ze^{i\alpha}|||G>||\left( \matrix {1 \cr 0 \cr 0 \cr . \cr .
\cr . \cr 0 } \right).
\end{eqnarray}
It has the isotropy (stationary) group $H=U(1) \times U(N-1)$
generated by the $H$-subset of the $\hat{\lambda}$-matrices.
Thereby, such representation of $Z|G>$ dictates the state-dependent
parametrization of the embedding isotropy group $H$ into $G=SU(N)$
and this parametrization is unstable relative general unitary
transformations in the following sense. Any transformation from $H$
leaves the $Z|G>$ intact, but arbitrary transformation from the
coset $G/H=U(N)/U(1) \times U(N-1)=CP(N-1)$ deforms the structure of
the $Z|G>$. It is easy to see if one applies the general coset
operator
\begin{eqnarray}
& \hat{T}(\tau,g) = \cr & \left( \matrix{\cos g\tau
&\frac{-p^{1*}}{g} \sin g\tau &\frac{-p^{2*}}{g}\sin g\tau
&.&\frac{-p^{N-1*}}{g}\sin g\tau \cr \frac{p^1}{g} \sin g\tau
&1+[\frac{|p^1|}{g}]^2 (\cos g\tau -1)&[\frac{p^1 p^{2*}}{g}]^2
(\cos g\tau -1)&.&[\frac{p^1 p^{N-1*}}{g}]^2 (\cos g\tau -1) \cr
.&.&.&.&. \cr .&.&.&.&. \cr .&.&.&.&. \cr \frac{p^{N-1}}{g}\sin
g\tau &[\frac{p^{1*} p^{N-1}}{g}]^2 (\cos g\tau
-1)&.&.&1+[\frac{|p^{N-1}|}{g}]^2 (\cos g\tau -1)} \right),
\end{eqnarray}
where $g=\sqrt{|p^1|^2+,...,+|p^{N-1}|^2}$ which effectively changes
this state dragging it along one of the geodesics in $CP(N-1)$
\cite{Le2}. In order to return to the (\ref{vac})-form one should
use $H$ transformations. Therefore the parameterizations of $H$ and
$G/H=CP(N-1)$ and their action are state-dependent (local in the
state space). It means that the Cartan's decomposition and the
representation of the $SU(N)$ generators should be state-dependent
\cite{Le2}. These local dynamical variables corresponding to the
$SU(N)$ group should be expressed now in terms of the local
coordinates $\pi^k_{(j)}=\frac{g^k}{g^j}$. Hence local dynamical
variables, their norms should be state-dependent, i.e. they are
functions of the local coordinates $\pi^k_{(j)}=\frac{g^k}{g^j}$ in
the complex projective Hilbert state space \cite{Le2} with the
Fubini-Study metric
\begin{equation}
G_{ik^*} = [(1+ \sum |\pi^s|^2) \delta_{ik}- \pi^{i^*} \pi^k](1+
\sum |\pi^s|^2)^{-2} \label{FS}
\end{equation}
and the affine connection
\begin{eqnarray}
\Gamma^i_{mn} = \frac{1}{2}G^{ip^*} (\frac{\partial
G_{mp^*}}{\partial \pi^n} + \frac{\partial G_{p^*n}}{\partial
\pi^m}) = -  \frac{\delta^i_m \pi^{n^*} + \delta^i_n \pi^{m^*}}{1+
\sum |\pi^s|^2} \label{Gamma}.
\end{eqnarray}

These generators give the infinitesimal shift of the $i$-component
of the generalized coherent state driven by the $\sigma$-component
of the unitary field $\Omega^{\sigma}$ rotating the standard Pauli,
Gell-Mann, ..., $\hat{\lambda}$-matrices of the $Alg SU(N)$
realization. They are defined as follows:
\begin{equation}
\Phi_{\sigma}^i = \lim_{\epsilon \to 0} \epsilon^{-1}
\biggl\{\frac{[\exp(i\epsilon \hat{\lambda}_{\sigma})]_m^i
g^m}{[\exp(i \epsilon \hat{\lambda}_{\sigma})]_m^j g^m
}-\frac{g^i}{g^j} \biggr\}= \lim_{\epsilon \to 0} \epsilon^{-1} \{
\pi^i_{(j)}(\epsilon \hat{\lambda}_{\sigma}) -\pi^i_{(j)} \}.
\end{equation}
Thereby, $\Phi^i_{\sigma}, \quad 1 \le \sigma \le N^2-1 $ are the
coefficient functions of the generators of the non-linear $SU(N)$
realization by the tangent vector fields to $CP(N-1)$.

These local dynamical variables (LDV's) realize a non-linear
representation of the unitary global $SU(N)$ group in the Hilbert
state space $C^N$. Namely, $N^2-1$ generators of $G = SU(N)$ may be
divided in accordance with the Cartan decomposition: $[H,H] \in H,
[B,H] \in B, [B,B] \in H$. The $(N-1)^2$ generators
\begin{eqnarray}
\Phi_h^i \frac{\partial}{\partial \pi^i}+c.c. \in H,\quad 1 \le h
\le (N-1)^2
\end{eqnarray}
of the isotropy group $H = U(1)\times U(N-1)$ of the ray (Cartan
sub-algebra) and $2(N-1)$ generators
\begin{eqnarray}
\Phi_b^i \frac{\partial}{\partial \pi^i} + c.c. \in B, \quad 1 \le b
\le 2(N-1)
\end{eqnarray}
are the coset $G/H = SU(N)/S[U(1) \times U(N-1)]$ generators
realizing the breakdown of the $G = SU(N)$ symmetry of the GCS.
Furthermore, the $(N-1)^2$ generators of the Cartan sub-algebra may
be divided into the two sets of operators: $1 \le c \le N-1$ ($N-1$
is the rank of $Alg SU(N)$) Abelian operators, and $1 \le q \le
(N-1)(N-2)$ non-Abelian operators corresponding to the
non-commutative part of the Cartan sub-algebra of the isotropy
(gauge) group.

Note, this representation of the generators has been found by the
infinitesimal action of the group parameters (unitary fields driven
by the infinitesimal real $\epsilon$). However the finite unitary
fields obey some fields equations expressing {\it the conservation
laws of the identity} which will be discussed below.

\section{Geometry of the quantum evolution}
Let me return to the geometry of the quantum evolution. For this aim
it is convenient to use well known Berry works.

Berry in his construction of the geometric phase used some intuitive
analogy between the tangent vector $\vec{e}$ to the some sphere
$S^2$ and the state vector $|\phi (X)>=|\phi (x_1,...,x_p)>$ whose
time dependence is generated by the periodic Hamiltonian
$\hat{H}(X(t)):X(T)=X(0)$. I will assume that vector state is a
tangent vector to the projective Hilbert space $CP(N-1)$ where
instead of an arbitrary parameters $X= (x_1,...,x_p)$ of the
Hamiltonian I will use intrinsic local projective coordinates
$\pi^k_{(j)}=\frac{g^k}{g^j}$ of the quantum states and local
dynamical variables. Their parallel transport is required to be in
agreement with the Fubini-Study metric. Then the affine connection
takes the place of the gauge potential of the non-Abelian type
playing the role of the covariant instant renormalization of the
dynamical variables during general transformations of the quantum
self-reference frame \cite{Berry164}.

Let me assume that $|G>=\sum_{a=0}^{N-1} g^a|a\hbar>$ is a ``ground
state" of some the least action problem, where for $a=0$ one has
\begin{eqnarray}
g^0(\pi^1_{j(p)},...,\pi^{N-1}_{j(p)})=R^2(R^2+
\sum_{s=1}^{N-1}|\pi^s_{j(p)}|^2)^{-1/2}
\end{eqnarray}
and for $a: 1\leq a = i \leq N-1$ one has
\begin{eqnarray}
g^i(\pi^1_{j(p)},...,\pi^{N-1}_{j(p)})=R \pi^i_{j(p)}(R^2+
\sum_{s=1}^{N-1}|\pi^s_{j(p)}|^2)^{-1/2}.
\end{eqnarray}
Then the velocity of the ground state evolution relative ``world
time" $\tau$ is given by the formula
\begin{eqnarray}\label{41}
|\Psi> \equiv |T> =\frac{d|G>}{d\tau}=\frac{\partial g^a}{\partial
\pi^i}\frac{d\pi^i}{d\tau}|\hbar a>+\frac{\partial g^a}{\partial
\pi^{*i}}\frac{d\pi^{*i}}{d\tau}|\hbar a> \cr
=|T_i>\frac{d\pi^i}{d\tau}+|T_{*i}>\frac{d\pi^{*i}}{d\tau}=H^i|T_i>+H^{*i}|T_{*i}>,
\end{eqnarray}
is the tangent vector to the evolution curve $\pi^i=\pi^i(\tau)$,
where
\begin{eqnarray}\label{42}
|T_i> = \frac{\partial g^a}{\partial \pi^i}|\hbar a>=T^a_i|\hbar a>,
\quad |T_{*i}> = \frac{\partial g^a}{\partial \pi^{*i}}|\hbar
a>=T^a_{*i}|\hbar a>.
\end{eqnarray}
I should emphasize that ``world time" is the  time of evolution from
the one GCS to another one which is physically distinguishable.
Thereby the unitary evolution of the action amplitudes generated by
(12) leads in general to the non-unitary evolution of the tangent
vector to $CP(N-1)$ associated with ``state vector" $|\Psi>$.

Then the variation velocity of the $|\Psi>$ is given by the equation
\begin{eqnarray}\label{43}
|A> &=&\frac{d|\Psi>}{d\tau} \cr &=&
(B_{ik}H^i\frac{d\pi^k}{d\tau}+B_{ik^*}H^i\frac{d\pi^{k*}}{d\tau}
+B_{i^*k}H^{i^*}\frac{d\pi^k}{d\tau} +B_{i^*
k^*}H^{i^*}\frac{d\pi^{k*}}{d\tau})|N>\cr &+&
(\frac{dH^s}{d\tau}+\Gamma_{ik}^s
H^i\frac{d\pi^k}{d\tau})|T_s>+(\frac{dH^{s*}}{d\tau}+\Gamma_{i^*k^*}^{s*}
H^{i*}\frac{d\pi^{k*}}{d\tau})|T_{s*}>,
\end{eqnarray}
where I introduce the matrix $\tilde{B}$ of the second quadratic
form whose components are defined by following equations
\begin{eqnarray}\label{45}
B_{ik}|N> =\frac{\partial |T_i>}{\partial \pi^k}-\Gamma_{ik}^s|T_s>,
\quad B_{ik^*}|N> = \frac{\partial |T_i>}{\partial \pi^{k*}} \cr
B_{i^*k}|N> =\frac{\partial |T_{i*}>}{\partial \pi^k}, \quad B_{i^*
k^*}|N> = \frac{\partial |T_{i*}>}{\partial
\pi^{k*}}-\Gamma_{i^*k^*}^{s*}|T_{s*}>
\end{eqnarray}
through the state $|N>$ normal to the ``hypersurface'' of the ground
states. Assuming that the ``acceleration'' $|A>$ is gotten by the
action of some linear ``Hamiltonian" $\hat{L}$ describing the
evolution (or a measurement), one has the ``Schr\"odinger equation
of evolution"
\begin{eqnarray}\label{56}
\frac{d|\Psi>}{d\tau}&=&-i\hat{L}|\Psi> \cr
&=&(B_{ik}H^i\frac{d\pi^k}{d\tau}+B_{ik^*}H^i\frac{d\pi^{k*}}{d\tau}
+B_{i^*k}H^{i^*}\frac{d\pi^k}{d\tau} +B_{i^*
k^*}H^{i^*}\frac{d\pi^{k*}}{d\tau})|N> \cr &+&
(\frac{dH^s}{d\tau}+\Gamma_{ik}^s
H^i\frac{d\pi^k}{d\tau})|T_s>+(\frac{dH^{s*}}{d\tau}+\Gamma_{i^*k^*}^{s*}
H^{i*}\frac{d\pi^{k*}}{d\tau})|T_{s*}>.
\end{eqnarray}
This ``Hamiltonian" $\hat{L}$ is non-Hermitian and its expectation
values is as follows:
\begin{eqnarray}\label{57}
<N|\hat{L}|\Psi>&=&
i(B_{ik}H^i\frac{d\pi^k}{d\tau}+B_{ik^*}H^i\frac{d\pi^{k*}}{d\tau}
+B_{i^*k}H^{i^*}\frac{d\pi^k}{d\tau} +B_{i^*
k^*}H^{i^*}\frac{d\pi^{k*}}{d\tau}),\cr <\Psi|\hat{L}|\Psi>&=&
iG_{p^*s}(\frac{dH^s}{d\tau}+\Gamma_{ik}^s
H^i\frac{d\pi^k}{d\tau})H^{p*}+iG_{ps^*}(\frac{dH^{s*}}{d\tau}+\Gamma_{i^*
k^*}^{s*} H^{i^*}\frac{d\pi^{k*}}{d\tau})H^p\cr
&=&i<\Psi|\frac{d}{d\tau}|\Psi>.
\end{eqnarray}
The minimization of the $|A>$ under the transition from point $\tau$
to $\tau+d\tau$ may be achieved by the annihilation of the
tangential component
\begin{equation}
\frac{dH^s}{d\tau}+\Gamma_{ik}^s H^i\frac{d\pi^k}{d\tau}=0, \quad
\frac{dH^{s*}}{d\tau}+\Gamma_{i^* k^*}^{s*}
H^{i^*}\frac{d\pi^{k*}}{d\tau}=0
\end{equation}
i.e. under the condition of the affine parallel transport of the
Hamiltonian vector field. The last equations in (25) shows that the
affine parallel transport of $H^i$ agrees with Fubini-Study metric
(13) leads to Berry's ``parallel transport" of $|\Psi>$ \cite{AS}.
Geometrically this picture corresponds to the special choice of the
moving reference frame
$\{|N>,|T_1>,...,|T_{N-1}>,|T_1*>,...,|T_{N-1}*>\}$ on $CP(N-1)$
that only ``longitudinal'' component along $|N>$ is alive.

The Berry's formula (1.24) \cite{Berry198} may be applied to the
eigen-vector with coordinates (18), (19) in the local coordinates
$\pi^i$. Supposing $R=1$ one found the 2-form as follows:
\begin{eqnarray}
V_{ik*}(\pi^i)= \Im \sum_{a=0}^{N-1} \{\frac{\partial
g^{a*}}{\partial \pi^i} \frac{\partial g^a}{\partial \pi^{k*}} -
\frac{\partial g^{a*}}{\partial \pi^{k*}} \frac{\partial
g^a}{\partial \pi^i} \} =\Im \sum_{a=0}^{N-1}\{T^a_i T^{a*}_k -
(T^a_k)^* T^{a}_i \} \cr = -\Im [(1+ \sum |\pi^s|^2) \delta_{ik}-
\pi^{i^*} \pi^k](1+ \sum |\pi^s|^2)^{-2}= - \Im G_{ik^*}
\label{form}
\end{eqnarray}
which closely related to Fubini-Study metric (quantum metric
tensor). There are, however, two important differences between
original Berry's formula referring to arbitrary parameters and this
2-form in local coordinates inherently connected with eigen-problem.

1. The $V_{ik*}(\pi^i)$ is the singular-free expression.

2. It does not contain two eigen-values, say, $E_n, E_m$ explicitly
, but implicitly $V_{ik*}$ depends locally on the choice of single
$\lambda_p$ through the dependence in local coordinates
$\pi^i_{j(p)}$.

Here arises a new problem (in comparison with the Berry's one) to
find field equations for the $SU(N)$ parameters $\Omega^{\alpha}$
leading to the affine parallel transport of the Hamiltonian field
$H^i=\Omega^{\alpha}\Phi^i_{\alpha}$. These field equations of
motion for quantum system whose `particles' do not exist a priory
but they are becoming during the evolution. But first of all we
should to introduce the notion of the ``dynamical space-time" which
is arises due to the natural evolution or the objective measurement
of some dynamical variable.

The ``probability" may be introduced now by pure geometric way like
$cos^2 \phi $ in tangent state space as follows.

For any two tangent vectors $D_1^i=<D_1|T_i>, D_2^i=<D_2|T_i>$ one
can define the scalar product
\begin{eqnarray}\label{}
(D_1,D_2)=\Re G_{ik^*} D_1^i D_2^{k^*}=\cos \phi_{1,2}
(D_1,D_1)^{1/2} (D_2,D_2)^{1/2}.
\end{eqnarray}
Then the value
\begin{eqnarray}\label{}
P_{1,2}(\pi^1_{j(p)},...,\pi^{N-1}_{j(p)})=\cos^2
\phi_{1,2}=\frac{(D_1,D_2)^2}{(D_1,D_1) (D_2,D_2)}
\end{eqnarray}
may be treated as a relative probability of the appearance of two
states arising during the measurements of two different dynamical
variables $D_1, D_2$ by the variation of the initial GCS
$(\pi^1_{j(p)},...,\pi^{N-1}_{j(p)})$.

\section{Evolution or Objective Measurement}
The $CP(N-1)$ manifold takes the place of the ``classical phase
space'' since its points, corresponding to the GCS, are most close
to classical states of motion. The interpretations may be given for
the points of $CP(N-1)$  as the ``Schr\"odinger's lump"
\cite{Penrose}. As we will see later, the important fact is in my
case the ``Schr\"odinger's lump" has the exact mathematical
description and clear physical interpretation: points of $CP(N-1)$
are the axis of the ellipsoid of the action operator $\hat{S}$, i.e.
extremals of the action functional $S[|F>]$. Then the velocities of
variation of these axis correspond to local Hamiltonian or different
local dynamical variables.

The basic content of $CP(N-1)$ points is their physical
interpretation as discriminators of physically distinguishable
quantum states. As such, they may be used as ``yes/no'' states of
some two-level ``detector". Let me assume that GCS described by
local coordinates $(\pi^1,...,\pi^{N-1})$ corresponds to the
original lump, and the coordinates $(\pi^1+\delta
\pi^1,...,\pi^{N-1}+\delta \pi^{N-1})$ correspond to the lump
displaced due to measurement. I will use a GCS
$(\pi^1_{j(p)},...,\pi^{N-1}_{j(p)})$ of some action operator
$\hat{S}=\hbar A(\hat{\eta^+}\hat{\eta})$ representing physically
distinguishable states. This means that any two points of $CP(N-1)$
define two ellipsoids differ at least by the orientations, if not by
the shape, as it was discussed above.

Local coordinates of the lump gives the a firm geometric tool for
the description of quantum dynamics during interaction which used
for a measuring process or evolution. The question that I would like
to raise is as follows: {\it what ``classical field'', i.e. field in
space-time, corresponds to the transition from the original to the
displaced lump?} In other words I would like to find the measurable
physical manifestation of the lump , which we shall call the ``field
shell", its space-time shape and its dynamics. The lump's dynamics
will be represented by (energy) frequencies distribution that are
not a priori given, but are defined by some field equations which
should established by means of variation problem applied to
operators represented by tangent vectors to $CP(N-1)$.

The eigen-problem $\hat{L}|\Psi>=\lambda|\Psi >$ for ``expectation
state" reads now as follows:
\begin{eqnarray}\label{43}
0&=&(B_{ik}H^i\frac{d\pi^k}{d\tau}+B_{ik^*}H^i\frac{d\pi^{k*}}{d\tau}
+B_{i^*k}H^{i^*}\frac{d\pi^k}{d\tau} +B_{i^*
k^*}H^{i^*}\frac{d\pi^{k*}}{d\tau})|N> \cr &+&
(\frac{dH^s}{d\tau}+\Gamma_{ik}^s H^i\frac{d\pi^k}{d\tau}+i\lambda
H^s)|T_s> \cr &+&(\frac{dH^{s*}}{d\tau}+\Gamma_{i^*k^*}^{s*}
H^{i*}\frac{d\pi^{k*}}{d\tau}+i\lambda H^{s*})|T_{s*}>.
\end{eqnarray}
Putting $\Gamma_{ik}^s=0$ one has simple solutions $H^s=C^s \exp
(-i\lambda \tau)$  restricting velocities $\frac{d\pi^k}{d\tau}$
through the equation
\begin{eqnarray}\label{49}
B_{ik}H^i\frac{d\pi^k}{d\tau}+B_{ik^*}H^i\frac{d\pi^{k*}}{d\tau}
+B_{i^*k}H^{i^*}\frac{d\pi^k}{d\tau} +B_{i^*
k^*}H^{i^*}\frac{d\pi^{k*}}{d\tau}=0.
\end{eqnarray}
The general self-consistent case is of course very complicated.

In order to build the spinor $\eta$ of the quantum question
$\hat{Q}$ \cite{Le6} in the local basis
$\{|N>,|T_1>,...,|T_{N-1}>,|T_1*>,...,|T_{N-1}*>\}$ for the
measurement of the ``Hamiltonian" $\hat{L}$ at corresponding GCS I
will use following equations
\begin{eqnarray}\label{513}
\alpha_{(\pi^1,...,\pi^{N-1})}=\frac{<N|\hat{L}|\Psi>}{<N|N>} \cr
\beta_{(\pi^1,...,\pi^{N-1})}=\frac{<\Psi|\hat{L}|\Psi>}
{<\Psi|\Psi>}.
\end{eqnarray}
Projector $\hat{Q}$ takes the place of dichotomic dynamical variable
(quantum question) for the discrimination of the self-energy of the
eigen-state at GCS (``vertical" motion along the normal state $|N>$)
and the perturbation energy that represents the velocity of
deformation of the GCS (``horizontal" motion along the tangent state
$|\Psi>$). Then from the infinitesimally close GCS
$(\pi^1+\delta^1,...,\pi^{N-1}+\delta^{N-1})$, whose shift is
induced by the interaction used for a measurement, one get a close
spinor $\eta+\delta \eta$ with the components
\begin{eqnarray}\label{514}
\alpha_{(\pi^1+\delta^1,...,\pi^{N-1}+\delta^{N-1})}=\frac{<N'|\hat{L}|\Psi>}{<N'|N'>}
\cr \beta_{(\pi^1+\delta^1,...,\pi^{N-1}+\delta^{N-1})}=
\frac{<\Psi'|\hat{L}|\Psi>}{<\Psi'|\Psi'>},
\end{eqnarray}
where the basis $(|N'>,|\Psi'>)$ is the lift of the parallel
transported $(|N>,|\Psi>)$ from the infinitesimally close point
$(\pi^1+\delta^1,...,\pi^{N-1}+\delta^{N-1})$ back to the
$(\pi^1,...,\pi^{N-1})$.

Now one should to find how the affine parallel transport connected
with the variation of coefficients $\Omega^{\alpha}$ in the
dynamical space-time associated with quantum question $\hat{Q}$.

The covariance relative transition from one GCS to another
\begin{eqnarray}
(\pi^1_{j(p)},...,\pi^{N-1}_{j(p)}) \rightarrow
(\pi^1_{j'(q)},...,\pi^{N-1}_{j'(q)})
\end{eqnarray}
and the covariant differentiation (relative Fubini-Study metric) of
vector fields provides the objective character of the ``quantum
question" $\hat{Q}$ and, hence, the quantum measurement. This serves
as a base for the construction of the dynamical space-time as it
will be shown below.

These two infinitesimally close spinors $\eta$ and $\eta+\delta
\eta$ may be expressed as functions of $\theta,\phi,\psi,R$ and
$\theta+\epsilon_1,\phi+\epsilon_2,\psi+\epsilon_3,R+\epsilon_4,$
and represented as follows
\begin{eqnarray}\label{s1}
\eta = R \left( \begin {array}{c} \cos \frac{\theta}{2}(\cos
\frac{\phi- \psi}{2} - i\sin \frac{\phi - \psi}{2}) \cr \sin
\frac{\theta}{2} (\cos \frac{\phi+\psi}{2} +i \sin
 \frac{\phi+\psi}{2})  \end {array}
 \right)
 = R\left( \begin {array}{c} C(c-is) \cr S( c_1+is_1)
\end
{array} \right)
\end{eqnarray}
and
\begin{eqnarray}
&\eta+\delta \eta = R\left( \begin {array}{c} C(c-is) \cr S(
c_1+is_1) \end {array} \right) \cr + & R\left( \begin {array}{c}
S(is-c)\epsilon_1-C(s+i c)\epsilon_2+
C(s+ic)\epsilon_3+C(c-is)\frac{\epsilon_4}{R} \cr
 C(c_1+is_1)\epsilon_1+S(ic_1-s_1)\epsilon_2-S(s_1-ic_1)\epsilon_3
+S(c_1+is_1)\frac{\epsilon_4}{R}
\end
{array}
 \right)
\end{eqnarray}
may be connected with infinitesimal ``Lorentz spin transformations
matrix'' \cite{G}
\begin{eqnarray}
L=\left( \begin {array}{cc} 1-\frac{i}{2}\tau ( \omega_3+ia_3 )
&-\frac{i}{2}\tau ( \omega_1+ia_1 -i ( \omega_2+ia_2)) \cr
-\frac{i}{2}\tau
 ( \omega_1+ia_1+i ( \omega_2+ia_2))
 &1-\frac{i}{2}\tau( -\omega_3-ia_3)
\end {array} \right).
\end{eqnarray}
Then accelerations $a_1,a_2,a_3$ and angle velocities $\omega_1,
\omega_2, \omega_3$ may be found in the linear approximation from
the equation
\begin{eqnarray}\label{equ}
\eta+\delta \eta = L \eta
\end{eqnarray}
as functions of the spinor components of the quantum question
depending on local coordinates $(\pi^1,...,\pi^{N-1})$.

Hence the infinitesimal Lorentz transformations define small
``space-time'' coordinates variations. It is convenient to take
Lorentz transformations in the following form $ct'=ct+(\vec{x}
\vec{a}) d\tau, \quad \vec{x'}=\vec{x}+ct\vec{a} d\tau
+(\vec{\omega} \times \vec{x}) d\tau$, where I put
$\vec{a}=(a_1/c,a_2/c,a_3/c), \quad
\vec{\omega}=(\omega_1,\omega_2,\omega_3)$ \cite{G} in order to have
for $\tau$ the physical dimension of time. The expression for the
``4-velocity" $ v^{\mu}$ is as follows
\begin{equation}
v^{\mu}=\frac{\delta x^{\mu}}{\delta \tau} = (\vec{x} \vec{a},
ct\vec{a}  +\vec{\omega} \times \vec{x}) .
\end{equation}
The coordinates $x^\mu$ of points in dynamical space-time serve in
fact merely for the parametrization of deformations of the ``field
shell'' arising under its motion according to non-linear field
equations \cite{Le3,Le4,Le5,Le6}.

\section{Field equations in the dynamical space-time}
The energetic packet  - ``particle'' associated with the ``field
shell'' is now described locally by the Hamiltonian vector field
$\vec{H}=\hbar
\Omega^{\alpha}\Phi^i_{\alpha}\frac{\partial}{\partial \pi^i} + c.c
$. Our aim is to find the wave equations for $\Omega^{\alpha}$ in
the dynamical space-time intrinsically connected with the objective
quantum measurement (evolution).

At each point $(\pi^1,...,\pi^{N-1})$ of the $CP(N-1)$ one has an
``expectation value'' of the $\vec{H}$ defined by a measuring
device. But a displaced GCS may by reached along one of the
continuum paths. Therefore the comparison of two vector fields and
their ``expectation values'' at neighboring points requires some
natural rule. The comparison makes sense only for the same
``particle'' represented by its ``field shell'' along some path. For
this reason one should have an identification procedure. The affine
parallel transport in $CP(N-1)$ of vector fields is a natural and
the simplest rule for the comparison of corresponding ``field
shells''.

The dynamical space-time coordinates $x^{\mu}$ may be introduced as
the state-dependent quantities, transforming in accordance with the
functionally local Lorentz transformations $\delta x^{\mu} = v^{\nu}
\delta \tau $ depend on the transformations of local reference frame
in $CP(N-1)$ as it was described in the previous paragraph.

Let us discuss now the self-consistent problem
\begin{equation}
v^{\mu} \frac{\partial \Omega^{\alpha}}{\partial x^{\mu} } = -
(\Gamma^m_{mn} \Phi_{\beta}^n+\frac{\partial
\Phi_{\beta}^n}{\partial \pi^n}) \Omega^{\alpha}\Omega^{\beta},
\quad \frac{d\pi^k}{d\tau}= \Phi_{\beta}^k \Omega^{\beta}
\end{equation}
arising under the condition of the affine parallel transport
\begin{eqnarray}
\frac{\delta H^k}{\delta \tau} &= &\hbar \frac{\delta
(\Phi^k_{\alpha} \Omega^{\alpha})}{\delta \tau}=0
\end{eqnarray}
of the Hamiltonian field. I will discuss the simplest case of
$CP(1)$ dynamics when $1\leq \alpha,\beta \leq3,\quad i,k,n=1$. This
system in the case of the spherical symmetry being split into the
real and imaginary parts takes the form
\begin{eqnarray}
\matrix{ (r/c)\omega_t+ct\omega_r=-2\omega \gamma F(u,v), \cr
(r/c)\gamma_t+ct\gamma_r=(\omega^2 - \gamma^2) F(u,v), \cr u_t=\beta
U(u,v,\omega,\gamma), \cr v_t=\beta V(u,v,\omega,\gamma), }
\label{self_sys}
\end{eqnarray}

It is impossible of course to solve this self-consistent problem
analytically even in this simplest case of the two state system, but
it is reasonable to develop a numerical approximation in the
vicinity of the following exact solution. Let me put $\omega=\rho
\cos \psi, \quad \gamma=\rho \sin \psi$, then, assuming for
simplicity that $\omega^2+\gamma^2=\rho^2=constant$, the two first
PDE's may be rewritten as follows:
\begin{equation}
\frac{r}{c}\psi_t+ct\psi_r=F(u,v) \rho \cos \psi.
\end{equation}
The one of the exact solutions of this quasi-linear PDE is
\begin{equation}
\psi_{exact}(t,r)=\arctan \frac{\exp(2c\rho F(u,v)
f(r^2-c^2t^2))(ct+r)^{2F(u,v)}-1}{\exp(2c\rho F(u,v)
f(r^2-c^2t^2))(ct+r)^{2F(u,v)}+1}, \label{ex_sol}
\end{equation}
where $f(r^2-c^2t^2)$ is an arbitrary function of the interval.

In order to keep physical interpretation of these equations I will
find the stationary solution for (43). Let me put $\xi=r-ct$. Then
one will get ordinary differential equation
\begin{equation}
\frac{d\Psi(\xi)}{d \xi} = -F(u,v) \rho \frac{\cos \Psi(\xi)}{\xi}.
\end{equation}
Two solutions
\begin{equation}
\Psi(\xi) =arctan(\frac{\xi^{-2M} e^{-2CM}-1}{\xi^{-2M} e^{-2CM}+1},
\frac{2\xi^{-M} e^{-2CM}}{\xi^{-2M} e^{-2CM}-1}  ),
\end{equation}
where $M=F(u,v) \rho$ are concentrated in the vicinity of the
light-cone look like solitary waves, see Fig.1.

\begin{figure}
\includegraphics[width=5in]{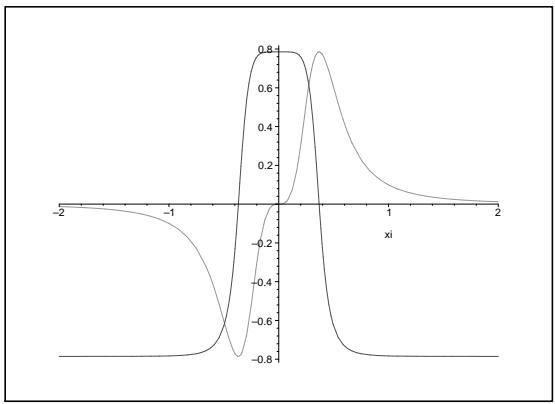}
\caption{Two solutions of (43) in the light-cone vicinity.}
\label{fig.1}
\end{figure}

Hence one may treat them as some ``potentials" for the local
coordinates of GCS $(u=\Re \pi^1,v=\Im \pi^1)$. The character of
these solutions should be discussed elsewhere.

\pagebreak {\bf Conclusion}

1. My purpose was to summarize in this work my last results
published presently in e-arXiv. We could not treat, of course, these
results as finally established. But I think that the natural
character of the assumptions which I put in the base of this theory
and the mathematically simple method of derivation of the
fundamental ``field shell" equations, will be attractive for future
investigations.

2. The concept of ``super-relativity" \cite{Le2,Le6,Le7,Le8} is in
fact some kind of ``hybridization" of internal and space-time
symmetries. The distinction in kind between SUSY where a priori
exists the extended space-time - ``super-space" and my approach is
that dynamical space-time arises under ``yes/no" objective quantum
measurement. This ``measurement" is represented by the covariant
differentiation of $SU(N)$ local dynamical variables in the
functional quantum phase space $CP(N-1)$ along the curve of
evolution.

\end{document}